\documentclass[aps,prstab,twocolumn,groupedaddress,showpacs]{revtex4}
\usepackage{graphicx}
\usepackage{bm}

\begin{document}
\title{Production of Enhanced Beam Halos via Collective Modes and Colored Noise}
\author{Ioannis V. Sideris$^1$ and Courtlandt L. Bohn$^2$}   
%\email{sideris@nicadd.niu.edu}
\affiliation{$^{1,2}$Department of Physics, Northern Illinois University, DeKalb, IL 60115\\
$^2$Fermi National Accelerator Laboratory, Batavia, IL 60510}

\begin{abstract}
We investigate how collective modes and colored noise conspire to produce a beam halo with much larger amplitude than could be generated by either phenomenon separately.
The collective modes are lowest-order radial eigenmodes calculated self-consistently for a configuration corresponding to a direct-current, cylindrically symmetric, warm-fluid Kapchinskij-Vladimirskij equilibrium.
The colored noise arises from unavoidable machine errors and influences the internal space-charge force.
Its presence quickly launches statistically rare particles to ever-growing amplitudes by continually kicking them back into phase with the collective-mode oscillations.
The halo amplitude is essentially the same for purely radial orbits as for orbits that are initially purely azimuthal; orbital angular momentum has no statistically significant impact.
Factors that do have an impact include the amplitudes of the collective modes and the strength and autocorrelation time of the colored noise.
The underlying dynamics ensues because the noise breaks the Kolmogorov-Arnol'd-Moser tori that otherwise would confine the beam.
These tori are fragile; even very weak noise will eventually break them, though the time scale for their disintegration depends on the noise strength.
Both collective modes and noise are therefore centrally important to the dynamics of halo formation in real beams.
\end{abstract}

\maketitle

\section{Introduction}
\label{sec:intro}
We recently demonstrated~\cite{bs03} that the combination of colored noise and global oscillations in intense charged-particle beams can create much larger halo amplitudes than would arise in the absence of noise.
This was done using generic `particle-core' models as representations of time-dependent potentials associated with nonequilibrium beams~\cite{chen91,gluckstern94}; the `core' established a time dependence in the form of a harmonic oscillation reminiscent of the presence of a global collective mode, and test particles orbited in response to that potential.
Ever-growing halos were found to form despite the fact that large-amplitude orbits spend considerable time under the influence of the external focusing forces, and the frequencies associated with these forces differ from those associated with the core oscillation, a circumstance that impedes resonance.
Thus, the noise has a key influence, boosting statistically rare particles to ever-growing amplitudes by continually kicking them back into phase with the core oscillation.
The importance of this finding lies in the accelerator's extreme sensitivity to beam loss.
For example, in a light-ion accelerator, beam impingement of just $\sim$1 W/m at energies exceeding $\sim$20 MeV will cause enough radioactivation to preclude hands-on machine maintenance~\cite{jameson96}.
In high-average-current machines, this amounts to just a few particles lost per meter, and large halos are thereby of practical concern, even if their outermost fringe is extremely tenuous.

Our previous analysis was restricted to radial orbits and centered on choosing the same initial conditions for all of the orbits.
Specifically, each orbit was assigned zero initial velocity and the same initial radius.
Because in a real beam each individual particle has its own distinct initial conditions (for example, the particles would start at different angular coordinates), each experiences its own manifestation of colored noise.
In other words, the noise was regarded to be spatially uncorrelated.
Thus, we sequentially computed 10,000 orbits while assigning to each orbit its own unique, random manifestation of the colored noise, and we cataloged the maximum amplitudes of these orbits.
Though this approach proved sufficient to demonstrate the noise-enhanced production of beam halo, it suffers a number of shortcomings.
First, it lacks self-consistency; with one exception, the oscillation frequencies of the core were chosen ad hoc, the exception relating to a space-charge-limited core.
Second, because only a single starting radius is sampled, it lacks the statistics of a full treatment; halo particles originating from, e.g., different radii are excluded.
Third, the contribution of nonradial orbits is likewise ignored.

The present paper offers a study that, by largely circumventing these shortcomings, is more thorough and systematic.
Herein we consider self-consistent collective oscillations in the context of a general framework.
Specifically, we consider a direct-current, cylindrically symmetric beam and model it as a warm-fluid Kapchinskij-Vladimirskij (KV) equilibrium configuration.
We then imagine the beam to be excited such that it possesses a self-consistent spectrum of collective, stable radial modes of oscillation as previously calculated by Lund and Davidson~\cite{lund98}.
The associated time-dependent space-charge force combines with the external focusing force to determine the equation of motion of test particles.
By populating the full configuration space with very many (typically $10^6$) test particles, assigning each test particle its own random manifestation of colored noise, and then tracking their orbits, we compute the evolution of the halo.
We do this for two extremes of initial particle velocities, the first corresponding once again to purely radial orbits, and the second corresponding to purely circular orbits.
The halo structure depends, of course, on (a) the beam parameters, which we combine into a single quantity, the space-charge tune depression, (b) the collective-mode parameters, specifically their amplitudes in that their frequencies are determined self-consistently, and (c) the noise parameters, specifically the noise strength and autocorrelation time.

In the investigation to follow, we quantify the various parametric dependencies.
Section~\ref{sec:methodology} explains our methodology in detail.
Section~\ref{sec:halo} presents an extensive array of results that together quantify how the beam and noise parameters conspire to produce large halos.
Included is an interpretation of the underlying dynamics in terms of the breaking of Kolmogorov-Arnol'd-Moser (KAM) tori~\cite{llbook} due to the presence of noise.
Section~\ref{sec:discussion} concludes by briefly summarizing the findings and, in view of them, identifying related phenomenology that will likely be inherent to fully self-consistent large $N$-body simulations of real beams.

\section{Methodology}
\label{sec:methodology}
As our foundation, we adopt directly the formalism of Strasburg and Davidson~\cite{strasburg}, hereafter called SD.
We consider an intense, direct-current charged-particle beam propagating in the $z$ direction at constant speed through a transport channel that imposes a constant, cylindrically symmetric, linear transverse focusing force.
The equilibrium beam is a warm-fluid Kapchinskij-Vladimirskij equilibrium, and collective modes are superposed upon this equilibrium.
These modes correspond to stable, axisymmetric flute perturbations and derive from linearizing the respective Vlasov-Maxwell-Poisson equations~\cite{lund98}.
The influence of the beam's self-fields on particle trajectories is properly included within the framework of the paraxial approximation.

We incorporate the beam parameters by way of the dimensionless self-field perveance $K$ given per the gaussian system of units as
\begin{equation}
K~=~\frac{2\rho q^2}{\beta^2\gamma^3mc^2},
\label{eq:perveance}
\end{equation}
wherein $\rho$ is the line density (number of particles per unit length), $q$ and $m$ are the particle charge and mass, respectively, $\beta$ and $\gamma$ are the usual relativistic factors, and $c$ is the speed of light.
The perveance then folds into the space-charge tune depression $\eta$ as
\begin{equation}
\label{eq:eta}
\eta~\equiv~\left[1-\left(\frac{\beta c}{\omega_fR_o}\right)^2K\right]^{1/2}\;,
\end{equation}
in which $R_o$ is the radius of the equilibrium beam and $\omega_f$ is the angular frequency associated with the bare external focusing force.
This parameter lies in the range $0\!\leq\!\eta\!\leq\!1$, the lower bound corresponding to the space-charge-limited beam, and the upper bound corresponding to zero space charge.

SD tabulate the potentials and frequencies corresponding to all of the axisymmetric flute modes.
These are collective normal modes, and as such are calculated using linear perturbation theory~\cite{lund98}.
The frequency of the $n^{th}$ such mode is given by
\begin{equation}
\label{eq:freq}
\omega_n(\eta^2)~=~\omega_f\sqrt{2[1+\eta^2(2n^2-1)]}\;.
\end{equation}
For their studies of particle dynamics, SD concentrate on the two lowest-order radial modes, $n\!=\!1,2$, and we shall do likewise.
We normalize the radial coordinate in terms of the radius $R_o$; however, unlike SD, we normalize time $t$ in terms of the angular frequency $\omega_f$, i.e., $t\!\rightarrow\!\omega_f t$.
In effect we are setting $R_o\!=\!1$ and $\omega_f\!=\!1$.

The axisymmetric flute modes are distinctly different from breathing modes.
The most elementary distinction is that the beam boundary is static ($R_o$ is constant) in the case of flute modes, but it oscillates in the case of breathing modes.
In both cases the beam is root-mean-square (rms) mismatched; however, for the flute modes the beam envelope is matched whereas for the breathing modes the envelope is mismatched.
We shall therefore use the terms ``envelope-matched'' and ``envelope-mismatched'' to refer to beams with axisymmetric flute modes and breathing modes, respectively.
For the warm-fluid KV beam, the equilibrium density profile exhibits a step-function discontinuity at the boundary.
In turn, the flute modes likewise include a discontinuity in the density profile at the boundary.
For example, consider the KV beam to be excited by the $n\!=\!1$ flute mode.
The density profile inside the beam is always uniform, but its magnitude oscillates.
To conserve particle number, this mode includes an oscillating surface charge, i.e., the density profile exhibits a Dirac delta function at the (stationary) envelope radius such that the integral over the beam volume is independent of time.
By contrast the lowest-order breathing mode entails a self-similar oscillation; the envelope radius oscillates, and the number density likewise oscillates but is everywhere uniform.

\subsection{Equation of Test-Particle Motion}
\label{subsec:equations}
To explore the dynamics of halo formation, we compute orbits of test particles that move in the total potential formed by the superposition of the external focusing potential and the space-charge potential.
The test particles contribute nothing to the total potential and do not interact with each other.
This means we treat the coarse-grained form of the beam's distribution function, thereby ignoring, e.g., discreteness effects from the individual point charges that comprise the beam.
Using the formalism herein, it may be possible to mimic discreteness effects by modeling them as appropriately weak gaussian white noise~\cite{kandrup03}, i.e., noise that has zero autocorrelation time, but we refrain from doing so in favor of concentrating on the influence of colored noise.

The equation of test-particle motion decomposes into two regimes, one for which the normalized radial coordinate $r\!<\!1$, and the other for which $r\!\geq\!1$.
If only the $n\!=\!1,2$ normal modes are excited, then the SD equation of test-particle motion with our normalization is
\begin{eqnarray}
\ddot{r}&+&\eta^2r-\frac{L^2}{r^3}-(1-\eta^2)r\lbrace\sqrt{\Gamma_1} \cos[\omega_1(\eta^2)t]\nonumber\\
&+&\sqrt{\Gamma_2}(1-\frac{3}{2}r^2)\cos[\omega_2(\eta^2)t]\rbrace=0\;\;\mbox{for $r\!<\!1$;}\nonumber\\
\ddot{r}&+&r-\frac{L^2}{r^3}-\frac{1-\eta^2}{r}=0\;\;\;\;\;\;\;\;\;\;\;\;\;\;\;\;\;\mbox{for $r\!\geq\!1$;}
\label{eq:motion}
\end{eqnarray}
in which $L$ is the dimensionless angular momentum, $\Gamma_n$ is the ratio of the rms electrostatic energy contained in collective mode $n$ to that contained in the equilibrium beam, and $\omega_n$ is given by Eq.~(\ref{eq:freq}) after setting $\omega_f\!=\!1$.
The constants $L$, $\Gamma_1$, and $\Gamma_2$ may be regarded as free parameters, the former in regard to the `geometry' of the test-particle orbit, and the latter in regard to the `amplitudes' of the respective collective modes.
However, because it derives from linear perturbation theory, for Eq.~(\ref{eq:motion}) to be valid, both $\Gamma_1$ and $\Gamma_2$ must be small compared to unity.
Note that the tune depression $\eta$ manifests itself not only in the mode frequency, but also in the frequency characterizing the effective focusing force acting on the test particle.
Hence, any noise that shows up in the tune depression influences {\it both} of these frequencies.

\subsection{Colored Noise}
\label{subsec:noise}
It is at this point that we depart in an important way from the SD treatment, for we wish to assess the extent to which noise, in combination with the collective mode(s), influences the particle dynamics.
This is a problem of practical importance; noise is unavoidable in real accelerators because they are imperfect.
Machine errors, as well as transitions, will feed space-charge fluctuations in that the beam evolves self-consistently in response to external influences.
Examples include forces from image charges due to irregularities in the accelerator hardware as well as radiofrequency and magnetic field errors, and in the lab frame the errors may themselves be time-independent or fluctuating due to, e.g., jitter~\cite{qiang01}.
>From the perspective of a beam particle, the effect of all of these machine imperfections is to impart time-dependent noise on the particle orbit, and thus we seek now to include this noise in the equation of test-particle motion.
Concerning our upcoming analysis, for zero noise we of course reproduce the dynamics that SD describe.
Thus, any differences that show up with nonzero noise are attributable solely to the presence of the noise itself.
Our main interest is to quantify how this noise influences the process of halo formation, and do so to an extent well beyond what we did previously.

Following the philosophy and procedure of our earlier investigation~\cite{bs03},  we add gaussian colored noise that samples an Ornstein-Uhlenbeck process~\cite{van}.
We do so in terms of a frequency fluctuation $\delta\omega(t)$.
Because the tune depression $\eta$ incorporates the space charge, we define this frequency fluctuation in terms of a fluctuating tune depression in a manner consistent with Eq.~(\ref{eq:freq}):
\begin{eqnarray}
\eta^2&\rightarrow&\eta^2+\delta\eta^2(t)~;\nonumber\\
\delta\eta^2(t)&\equiv&\omega_1\delta\omega(t)=\sqrt{2(1+\eta^2)}\delta\omega(t)\;.
\label{eq:noisyeta}
\end{eqnarray}
The frequency fluctuation $\delta\omega(t)$ henceforth represents the noise.
Thus, everywhere it occurs in the equation of motion, Eq.~(\ref{eq:motion}), the quantity $\eta^2$ is replaced by $\eta^2\!+\!\delta\eta^2(t)$, with $\delta\eta^2(t)$ given by the last expression in Eq.~(\ref{eq:noisyeta}) above.
Note, for example, that the noise will still manifest itself in Eq.~(\ref{eq:motion}) even if no collective mode is excited.

The first two moments of $\delta\omega(t)$ fully determine the statistical properties of the noise:
\begin{eqnarray}
\langle\delta\omega(t)\rangle&=&0~;\nonumber\\
\langle\delta\omega(t)\delta\omega(t_1)\rangle&=&A^2\exp(-|t-t_1|/t_c)\;;
\label{eq:noise}
\end{eqnarray}
in which $t_c$ denotes the autocorrelation time, i.e., the time scale over which the signal changes appreciably.
The special case of white noise corresponds to the limit $t_c\!\rightarrow\!0$.
After generating a colored-noise signal using an algorithm first presented in Ref.~\cite{pogorelov99}, we compute $|A|\leftrightarrow\langle|\delta\omega|\rangle$ which then constitutes the measure of noise strength.
Example manifestations of colored noise for various noise strengths and autocorrelation times are plotted in Fig.~1 of Ref.~\cite{bs03}.

The noise should be viewed from the perspective of the charged particle as it progresses along its trajectory.
A typical particle will respond to space-charge fluctuations that change over a time scale comparable to, e.g., a plasma period (which is thus a measure of the minimum autocorrelation time).
It will likewise respond to stochastic changes in the external fields arising as the particle transits the beam line.
Such changes can correspond to, e.g., the spacing between hardware components.
Accordingly a hierarchy of autocorrelation times comprises the actual noise a particle experiences.
In a real beam each individual particle will have its own distinct initial conditions and thus experience a manifestation of the noise differing from that seen by each of the other particles.
Hence, at each successive time step during an orbit integration, a randomly generated frequency fluctuation is computed in keeping with the specified statistical properties of the colored noise.
Then this frequency fluctuation is converted to a fluctuation in the space-charge tune depression per Eq.~(\ref{eq:noisyeta}), and the so-modified tune depression is inserted into the equation of motion, Eq.~(\ref{eq:motion}), whereby the fluctuation influences the next time step.
How the noise quantitatively affects halo formation depends on its strength and its autocorrelation time, dependencies that we quantify herein.

\subsection{Initial Distribution of Test-Particle Orbits}
\label{subsec:initial}
In keeping with the objective of retaining as much realism as possible, we choose a beam intermediate between the space-charge-limited beam ($\eta\!=\!0$) and the zero-space-charge beam ($\eta\!=\!1$).
Specifically, we keep $\eta$ fixed at the value $\eta\!=\!0.3$ for our entire investigation.

We choose an initial distribution of test particles that spans all of the dynamically interesting regions of configuration space and falls gradually to a low-density tail.
Of course there are numerous ways to do this; one is to choose a distribution corresponding to a configuration of thermal equilibrium (TE)~\cite{reiser,brown95}.
We construct a cylindrically symmetric TE configuration of test charges following a procedure recently used to devise spherically symmetric TE configurations~\cite{TE}.
The associated dimensionless Poisson equation is
\begin{equation}
\frac{1}{R}\frac{d}{dR}\bigg(R\frac{d\Phi}{dR}\bigg)=-e^{-\frac{1}{2}\Omega^2R^2-\Phi(R)}=-n(R)\;,
\label{eq:poisson}
\end{equation}
wherein $R$ is a dimensionless radial coordinate, $n(R)$ denotes the number density normalized to the central density, $\Omega$ is a dimensionless quantity governing the strength of the external focusing force vis-\`{a}-vis the collective space-charge force, and $\Phi(R)$ is the dimensionless space-charge potential.
For the value of $\Omega$ we choose $\Omega\!=\!(1\!+\!10^{-3.5}\!-\!10^{-9})/\sqrt{2}$.
Integrating Eq.~(\ref{eq:poisson}) numerically using this specific value of $\Omega$ yields a TE density profile corresponding to a tune depression $\eta\!\simeq\!0.3$~\cite{brown95}.

Length and time are normalized differently in Eq.~(\ref{eq:poisson}) than in Eq.~(\ref{eq:motion}).
In keeping with the desire to span all of the dynamically interesting regions of configuration space, we simply rescale the density distribution $n(R)$ calculated from Eq.~(\ref{eq:poisson}) so that its rms radius $\tilde{R}$ corresponds to the full radius $R_o$ of the warm-fluid KV distribution: $\tilde{R}\!=\!R_o\!=\!1$.
This clearly places a sizeable population of test particles, that corresponding to much of the density tail, outside the KV `core'.
It also mimics, e.g., an inference from the recent beam-halo experiment at the Low-Energy Demonstration Accelerator at Los Alamos National Laboratory that the input beam for this experiment carried a sizeable tail in its distribution~\cite{qiang02}.
For all of our investigations the initial radii of the test particles follows this distribution.
Most of our simulations involve $N\!=\!10^6$ test particles, a number sufficient to constitute a good statistical sample.
In principle, the tenuous tail of the density profile extends to infinity, but in practice there is a finite radius to the $N$-body representation of the density because $N$ is finite.

For most of our investigations the initial test-particle velocities are all set to zero, corresponding to purely radial orbits, in which case we then replace $r(t)$ by $x(t)$, and $r(t)\!<\!1$ or $\geq\!1$ by $|x(t)|\!<\!1$ or $\geq\!1$, respectively, in Eq.~(\ref{eq:motion}).
We also, however, consider another `limiting' case, that for which all the orbits are initially circular.
Given a radius $r_c$ of the initially circular orbit, the respective dimensionless angular momentum $L$, a quantity taken to be conserved, follows from Eq.~(\ref{eq:motion}):
\begin{equation}
L^2~=~\left\{\begin{array}{ll}
\eta^2r_c^4 &\mbox{for $r_c\!<\!1$,}\\
r_c^4-(1-\eta^2)r_c^2 &\mbox{for $r_c\!\geq\!1$.}
\end{array} \right.
\label{eq:L}
\end{equation}
As is shown and discussed in Sec.~\ref{subsubsec:circular} below, the influence of noise on circular orbits that start with $r_c\!<\!1$ is essentially the same as for the purely radial orbits.
Consequently, the halo population is similar for both cases.

\subsection{Orbit Integrations}
\label{sec:integration}
We integrate the equation of motion using a fifth-order Runge-Kutta algorithm with variable time step~\cite{NR} taking the initial time step to be 0.01 `differential-equation' (DE) units.
We evolve each orbit for a total time 512 DE units, which corresponds to 40-60 orbital periods depending on the initial conditions for the respective orbit.
Thus, for example, the total integration time is comparable to the transit time of the beam through a large proton linear accelerator such as that associated with the Spallation Neutron Source~\cite{sns}.
In the absence of time-dependence and noise, i.e., with $\Gamma_1\!=\!\Gamma_2\!=\!\langle|\delta\omega|\rangle\!=\!0$, the algorithm conserves energy within a fractional error $10^{-9}$ at each time step and within $10^{-7}$ over the whole integration.

Our investigation spans a broad sector of the parameter space in that the space-charge tune depression, set at $\eta\!=\!0.3$, is the only parameter that is never varied.
We treat all combinations of the following parametric values: mode amplitudes $\Gamma_{1,2}\!=\!0.05$, 0.10, and 0.20; noise strengths $\langle|\delta\omega|\rangle\!=\!0$, 0.001, 0.01, and 0.1; autocorrelation times $t_c\!=\!0.5$, 1, 2, 10, 80, and 160; and test-particle sample sizes $N\!=\!10^2$, $10^3$, $10^4$, $10^5$, and $10^6$.
Most of the plots shown herein pertain to the specific choice $t_c\!=\!80$; however, excursions to lower and higher values are included to provide a check on the sensitivity of halo formation to the autocorrelation time of the noise.
In a real machine, of course, the noise will incorporate a range, or `superposition', of autocorrelation times and strengths.
In addition, with one exception (Fig.~\ref{fig:momentum.ps}), all of the plots we show pertain to radial orbits, i.e., cases for which all of the test particles have $L\!=\!0$ in Eq.~(\ref{eq:motion}).

\section{Influence of Colored Noise on Halo Formation}
\label{sec:halo}
SD explored the dynamics of test-particle motion in the absence of noise.
The form of their equation of motion differed slightly from our Eq.~(\ref{eq:motion}) in that they normalized the time in terms of the space-charge-depressed focusing frequency rather than the external focusing frequency.
Notwithstanding the different normalization, the physical content remains unchanged.
SD discovered that the time-dependent potential associated with the presence of a collective axisymmetric flute mode, even if only weakly excited, establishes a chaotic region of phase space in the outer regions of the beam.
They also found that this feature is not present in the phase space of an envelope-mismatched beam having a similar level of rms mismatch but no collective mode (cf. Fig. 9 of Ref.~\cite{strasburg}).
An important consequence of the chaotic sea is that orbits entering it can stochastically explore a larger region of phase space, thereby gaining more energy and correspondingly larger orbital amplitude.
SD consequently demonstrated that the excitation of collective modes leads to a halo significantly larger than that generated by an envelope mismatch, and that the difference is due to destabilization of KAM surfaces by the collective mode.
As we will now show, the presence of colored noise substantially enhances the influence of the collective modes.

\subsection{Orbital Dynamics}
\label{subsec:orbits}
\begin{figure}
\includegraphics{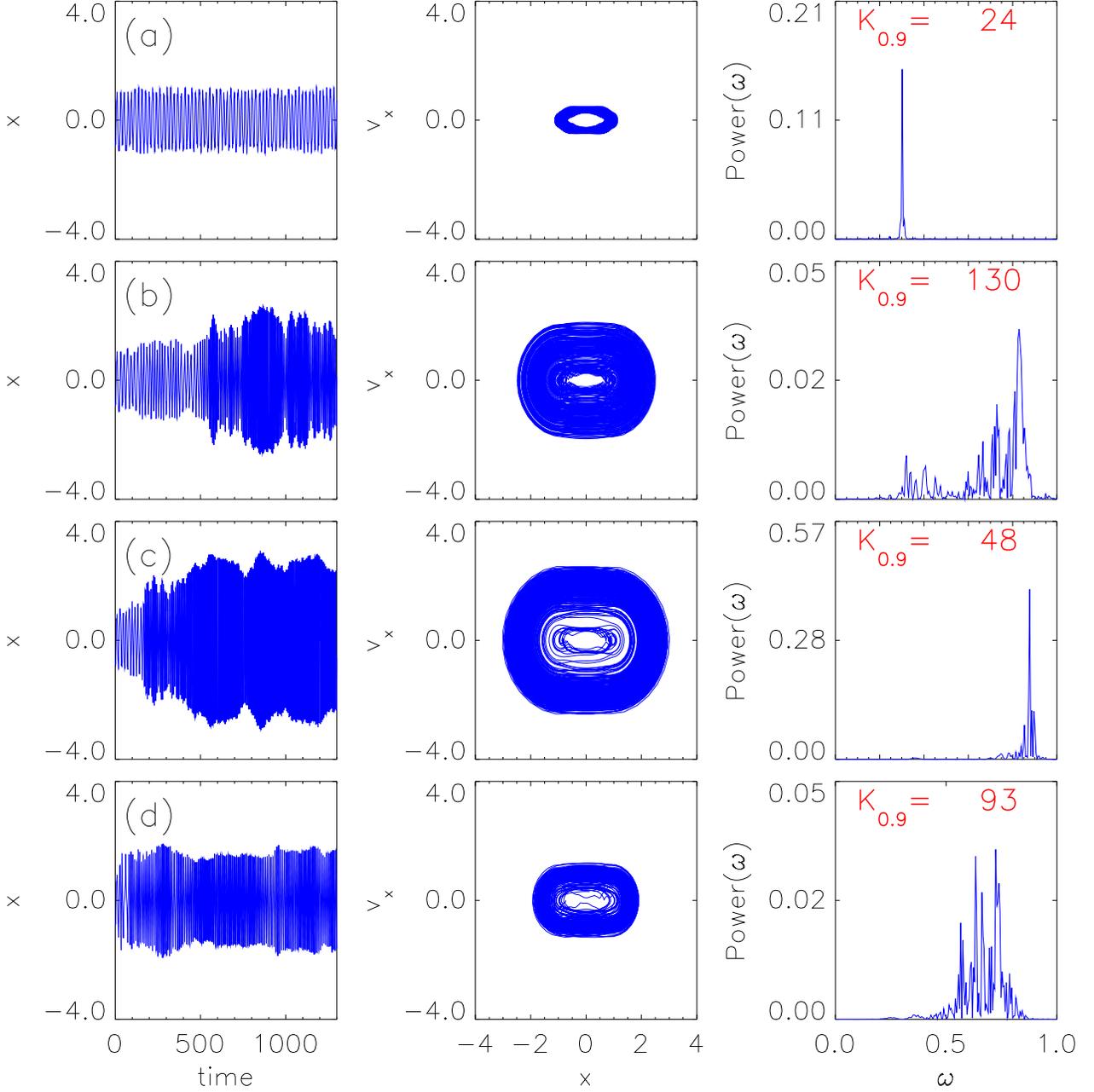}
\caption{\label{fig:spectrum.ps}
Plots of an example orbit having initial conditions $x(0)\!=\!-0.733407$, $\dot{x}\!=\!0$, in the presence of various noise strengths with $\Gamma_1\!=\!0.1$,  $\Gamma_2\!=\!0$, $t_c\!=\!80$.
The orbit is plotted in configuration space $x$ vs.~$t$ (left panel) and in phase space $\dot{x}$ vs.~$x$ (center panel), along with its corresponding power spectrum (right panel) wherein the complexity $K_{0.9}$ is provided as a measure of orbital chaoticity (see Sec.~\ref{subsec:orbits}).
The four rows correspond to different noise strengths: 
(a) $\langle|\delta\omega|\rangle\!=\!0$, 
(b) $\langle|\delta\omega|\rangle\!=\!0.001$, 
(c) $\langle|\delta\omega|\rangle\!=\!0.01$, and 
(d) $\langle|\delta\omega|\rangle\!=\!0.1$.}
\end{figure}

Halo formation is inextricably linked to the dynamics of individual orbits.
Consequently, a close inspection of what happens to an individual orbit because of the noise will be instructive.
We arbitrarily select an orbit that originates deep in the interior of the beam; the initial conditions are $x\!=\!-0.733407$ and $\dot{x}\!=\!0$.
We then integrate the orbit for 2048 DE units to obtain good frequency resolution in its power (Fourier) spectrum.
The trajectory and power spectrum of this orbit are plotted in Fig.~\ref{fig:spectrum.ps} with $\Gamma_1\!=\!0.1$, $\Gamma_2\!=\!0$, and $t_c\!=\!80$, and for a sequence of successively increasing noise strengths.
With zero noise the power spectrum is sharply peaked at a single frequency indicating that the orbit is periodic, hence regular, and the trajectory $x(t)$ clearly reflects this periodicity.
However, in the presence of even weak noise the orbit clearly becomes chaotic, having a power spectrum that features continua.
A useful measure of chaos is the number of frequencies $K_f$ that together contain a given fraction $f$ of the spectral power.
This measure is called the `complexity' of the orbit; a common choice is $f\!=\!0.9$~\cite{TE}.
Accordingly, a broader spectrum indicates a higher degree of chaos.
Concerning the orbit in Fig.~\ref{fig:spectrum.ps}, the noise has obviously placed it in a chaotic region of the phase space established as a consequence of the time-dependent potential associated with the collective mode.
How this happens is clarified in Sec.~\ref{subsec:tori} below.

As is also apparent from Fig.~\ref{fig:spectrum.ps}, the degree of orbital chaoticity as quantified via the complexity $K_{0.9}$ is not necessarily a simple, i.e., monotonic, function of the noise strength.
To reiterate, the power spectrum, hence the complexity, derives from the history of the orbit and thereby reflects a superposition of successive short-time behaviors.
An orbit that spends a relatively large fraction of time at large amplitudes, over which the net force is predominantly that of the harmonic external potential, will tend to be `more regular' and have smaller $K_{0.9}$.
As concerns a single, specific orbit (so no phase-space statistics are involved), what matters is not so much the amplitude of the noise, but rather whether a sequence of noise-induced kicks happens to make the orbit more chaotic, and these kicks are, of course, unpredictable {\it a priori}.
Our experiments indicate that a sequence of kicks leading to increased orbital chaoticity and/or increased orbital amplitude will occur sooner for some orbits and later for others.
In simulations involving many test particles distributed over a range of initial conditions, features of the evolving test-particle distribution are thus manifestly statistical.

\subsection{Evolution of the Halo Amplitude}
\label{subsec:noisyhalo}
We now evolve initial distributions of $N\!=\!10^6$ test particles constructed per the prescription of Sec.~\ref{subsec:initial}.
As the orbit integrations progress, we record a `snapshot' of the test-particle positions once every eight DE time units.
This interval approximately corresponds to the period of a typical orbit in the unperturbed SD potential, which we call the `dynamical time' $t_D$: 8 DE units $\simeq$1 $t_D$.
For every snapshot we record the largest radius reached by any of the $N$ particles; the collection of these radii represents the evolving halo amplitude $R_H(t)$.

\begin{figure}
\includegraphics{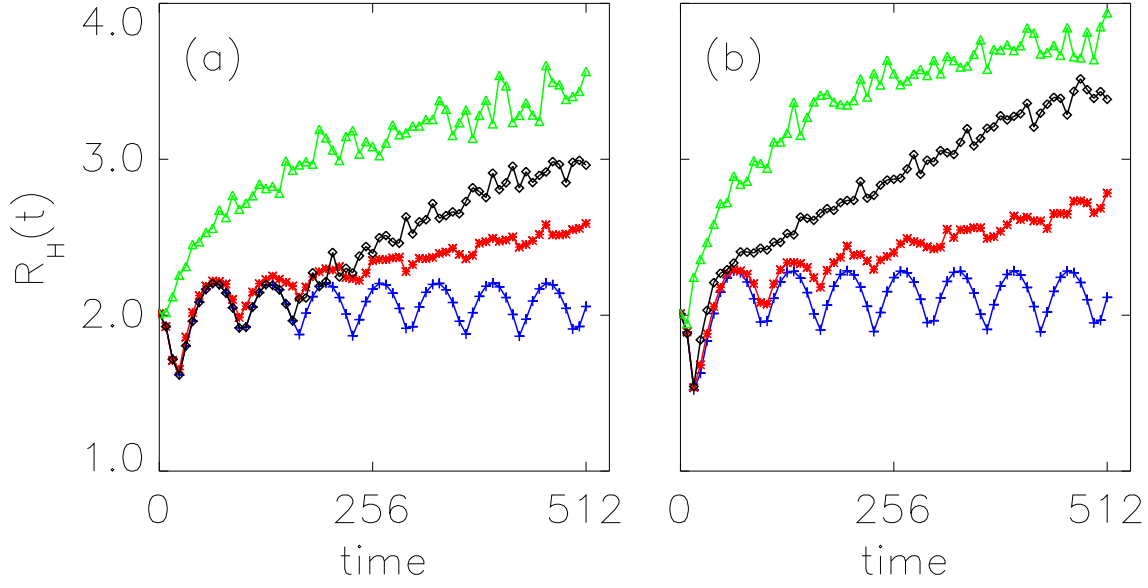}
\caption{\label{fig:R_H_t.ps}
Halo amplitude $R_H$ vs.~$t$ with $\Gamma_2\!=\!0$, $t_c\!=\!80$, and (a) $\Gamma_1\!=\!0.05$, (b) $\Gamma_1\!=\!0.1$.
The number of test particles is $N\!=\!10^6$.
The four curves correspond to four different noise amplitudes.
Blue curve with crosses: $\langle|\delta\omega|\rangle\!=\!0$.
Red curve with asterisks: $\langle|\delta\omega|\rangle\!=\!0.001$.
Black curve with diamonds: $\langle|\delta\omega|\rangle\!=\!0.01$.
Green curve with triangles: $\langle|\delta\omega|\rangle\!=\!0.1$.}
\end{figure}

Example results are plotted versus time in Fig.~\ref{fig:R_H_t.ps}, for which the mode amplitudes and autocorrelation time are fixed at $\Gamma_1\!=\!0.05$ or $0.1$, $\Gamma_2\!=\!0$, and $t_c\!=\!80$, and the noise strength  $\langle|\delta\omega|\rangle$ is varied from zero upward.
As the figure indicates, in the absence of noise the halo amplitude is quasiperiodic, and its time-averaged value stays the same, i.e., it does not grow.
This is as expected~\cite{gluckstern94,strasburg}.
A particle `resonantly' coupled to the collective mode is kicked to larger amplitudes.
However, because its orbital frequency changes as its amplitude changes, at sufficiently large amplitude the particle decouples from the mode and its amplitude ceases to grow.
Differences between the external focusing force and the collective space-charge force thus impose a hard upper bound on the halo amplitude.
The presence of noise, however, drastically changes this scenario.
Occasionally successive kicks from the noise will happen to be `just right' to keep a particle in phase with the mode for an effectively longer time and thereby push it beyond the upper bound (i.e., outer KAM torus) characterizing the noise-free case.
The halo amplitude $R_H(t)$ continues to grow, and the growth appears to be almost linear with time (at least after the first few oscillations).
Over the range of noise strengths $\langle|\delta\omega|\rangle$ we explore, both stronger noise and larger mode amplitudes enhance halo growth.
Moreover, when the noise is strong (e.g., $\langle|\delta\omega|\rangle\!=\!0.1$, i.e., roughly 10\% of the collective-mode frequency), pronounced halo growth occurs in just a few ($\sim$ 5) dynamical times.

\begin{figure}
\includegraphics{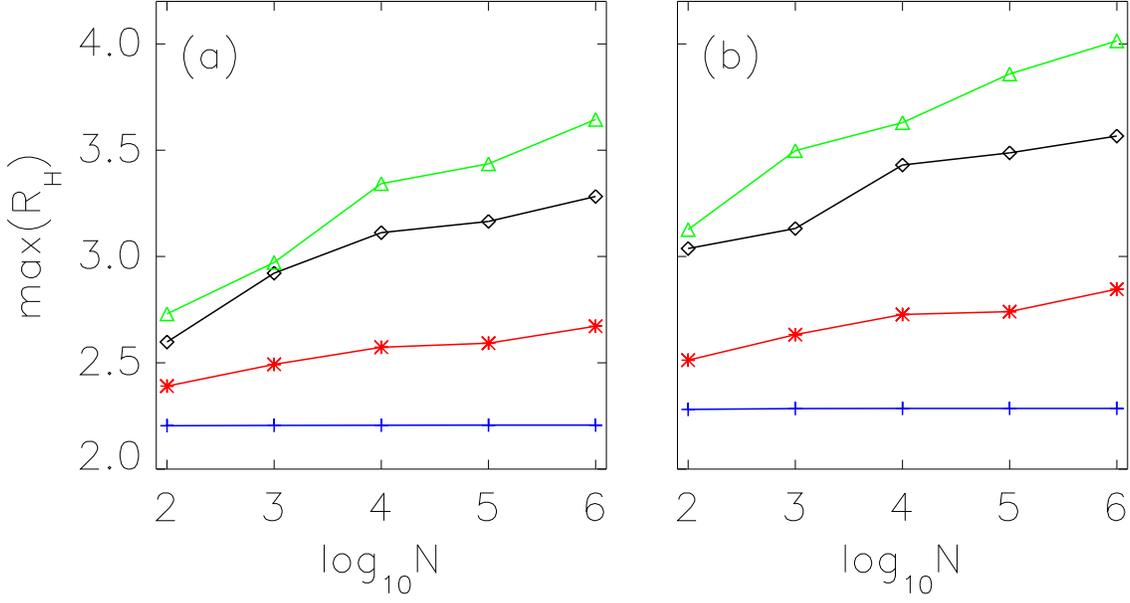}
\caption{\label{fig:max_R_H.ps}
Maximum halo amplitude $max(R_H)$ reached over a duration $t\!=\!512$ DE units vs.~the logarithm of the test-particle population $N$ with $\Gamma_2\!=\!0$ and $t_c\!=\!80$, and with (a) $\Gamma_1\!=\!0.05$, and (b) $\Gamma_1\!=\!0.1$.
The four curves correspond to four different noise amplitudes.
Blue curve with crosses: $\langle|\delta\omega|\rangle\!=\!0$.
Red curve with asterisks: $\langle|\delta\omega|\rangle\!=\!0.001$.
Black curve with diamonds: $\langle|\delta\omega|\rangle\!=\!0.01$.
Green curve with triangles: $\langle|\delta\omega|\rangle\!=\!0.1$.} 
\end{figure}

Consider the largest orbital amplitude reached by any particle over the course of a simulation, i.e., the largest halo amplitude, and denote this amplitude as $max(R_H)$.
This quantity will of course vary with the number of test particles $N$ in the simulation.
Because the number of particles that can be incorporated into $N$-body simulations is inherently limited by available computational power, it is of interest to know how sensitive the halo amplitude can be to the choice of $N$.
To quantify this sensitivity we adjust the test-particle population between $10^2\!\leq\!N\!\leq\!10^6$ particles.
Then we perform a number of experiments with different noise strengths and different values of $\Gamma_1$ (with $\Gamma_2\!=\!0$ and $t_c\!=\!80$).
Results pertaining to $\Gamma_1\!=\!0.05$ and $0.1$ appear in Fig.~\ref{fig:max_R_H.ps}.
Because there is a hard upper bound to the halo amplitude in the absence of noise, $max(R_H)$ is essentially independent of $N$ for the case $\langle|\delta\omega|\rangle\!=\!0$ provided $N$ is sensibly large.
This is not true when noise is present; the noise establishes a quasi-logarithmic dependence of $max(R_H)$ on $N$, a finding that is in keeping with our earlier results~\cite{bs03}.
Larger values of $\Gamma_1$ yield larger values of $max(R_H)$, but the scaling of $max(R_H)$ with $N$ remains roughly the same.

\begin{figure}
\includegraphics{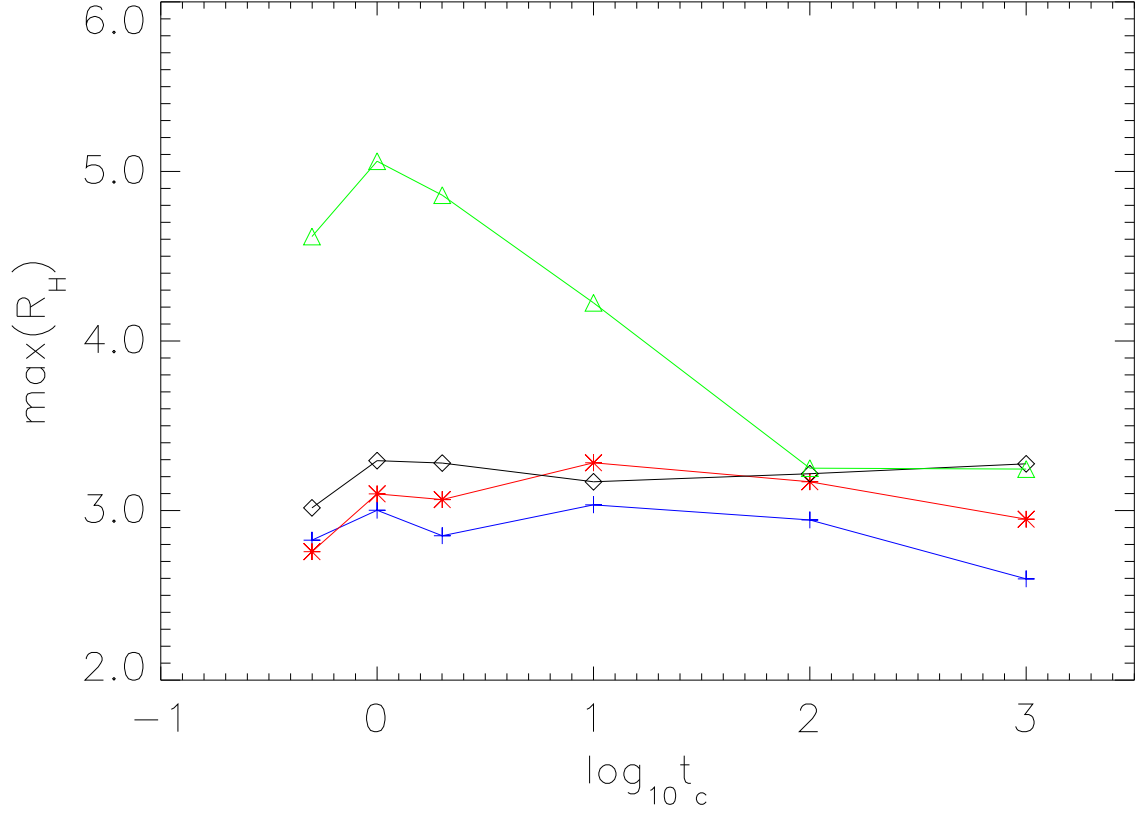}
\caption{\label{fig:tc.ps}
Maximum halo amplitude $max(R_H)$ vs.~the logarithm of the autocorrelation time $t_c$ computed for $N\!=\!10^4$ test particles with $\Gamma_1\!=\!0.05$ and $\Gamma_2\!=\!0$.
Blue curve with crosses: $\langle|\delta\omega|\rangle\!=\!0.002$.
Red curve with asterisks: $\langle|\delta\omega|\rangle\!=\!0.01$.
Black curve with diamonds: $\langle|\delta\omega|\rangle\!=\!0.03$.
Green curve with triangles: $\langle|\delta\omega|\rangle\!=\!0.1$.}
\end{figure}

Results presented thus far correspond to a single autocorrelation time $t_c\!=\!80\!\simeq\!10\;t_D$.
What happens if $t_c$ is much shorter or much longer?
Plots of maximum halo amplitude $max(R_H)$ versus $t_c$ for a sequence of increasing noise strengths $\langle|\delta\omega|\rangle$, and with only the $n\!=\!1$ collective mode active, appear in Fig.~\ref{fig:tc.ps}.
Data points in these plots each correspond to a sample of $N=10^4$ test particles; simulations with $N\!=\!10^6$ are found to give similar results but, of course, they involve much longer run times.
In most cases the halo extent is seen to be only weakly dependent on autocorrelation time.
The exception pertains to large noise strength: $\langle|\delta\omega|\rangle\!=\!0.1$, i.e., a 10\% fluctuation in the collective-mode frequency, generates substantially larger halo for $t_c\!<\!100$.
The presence of such large noise would seem to be anomalous in a real accelerator, and one might thus presume the halo amplitude will normally be independent of hardware details associated with the establishment of noise correlations.
However, under circumstances that lead to a turbulent beam as might reside, for example, at and just downstream of the beam source and at large hardware transitions, one might indeed expect the particle orbits to experience large noise from space charge locked in the turbulent eddies.
Such circumstances would seem normally to be transient, with the large-scale turbulence mixing away in a few dynamical times.
Nevertheless, because it would form rapidly, a sizeable halo would likely arise as evidenced from the $\langle|\delta\omega|\rangle\!=\!0.1$ curves in Fig.~\ref{fig:R_H_t.ps}.

\subsection{Evolution of the Test-Particle Distribution}
\label{subsec:distribution}
\subsubsection{Halo Density}
\begin{figure}
\includegraphics{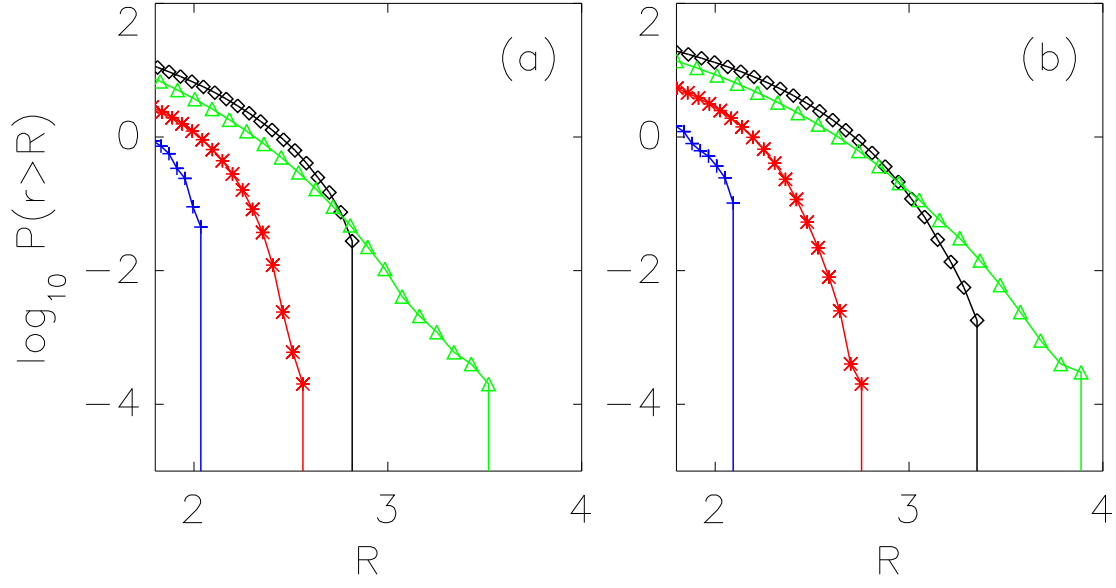}
\caption{\label{fig:log_P_r_R_r.ps}
Percentage $P(r\!>\!R)$ vs.~$R$ of test particles ($N\!=\!10^6$) lying outside radius $R$ at the end of the simulation ($t\!=\!512$) for various noise strengths with (a) $\Gamma_1\!=\!0.05$, and (b) $\Gamma_1\!=\!0.1$.
Fixed parameters are $t_c\!=\!80$ and $\Gamma_2\!=\!0$.
Blue curve with crosses: $\langle|\delta\omega|\rangle\!=\!0$.
Red curve with asterisks: $\langle|\delta\omega|\rangle\!=\!0.001$.
Black curve with diamonds: $\langle|\delta\omega|\rangle\!=\!0.01$.
Green curve with triangles: $\langle|\delta\omega|\rangle\!=\!0.1$.}
\end{figure}

Not just the amplitude of the halo is of interest, but also its density profile.
A convenient and meaningful representation of the test-particle distribution is obtained by calculating the percentage of particles lying outside a radius $R$; let us call this $P(r\!>\!R)$ while noting $P(r\!>\!R)\!\rightarrow\!100$\% as $R\!\rightarrow\!0$.
Plots of $\log_{10}[P(r\!>\!R)]$ versus $R$ computed at $t\!=\!512$ appear in Fig.~\ref{fig:log_P_r_R_r.ps}.
Here again, the mode amplitudes and autocorrelation time are fixed at $\Gamma_1\!=\!0.05$ or $0.1$, $\Gamma_2\!=\!0$, and $t_c\!=\!80$, and the noise strength $\langle|\delta\omega|\rangle$ is varied over a considerable range.
As a general trend the distribution spreads to larger radii, i.e., the halo amplitude grows, as the noise strength increases.

\begin{figure}

\includegraphics{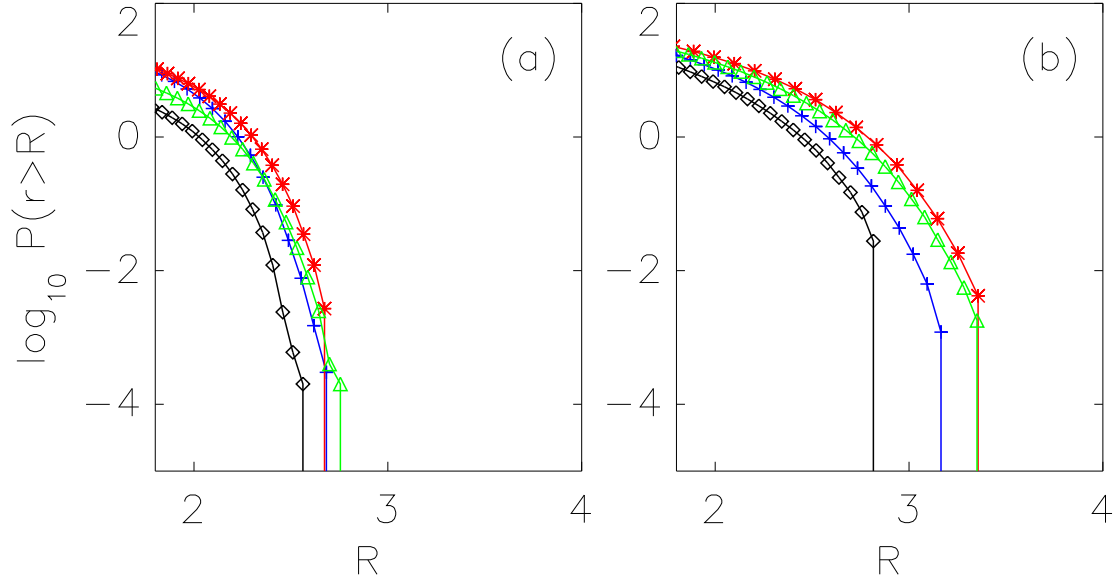}
\caption{\label{fig:log_P_r_R_r_2m.ps}
Plots of $P(r\!>\!R)$ vs.~$R$ in the presence of both the $n\!=\!1$ and $n\!=\!2$ modes ($N\!=\!10^6$) at the end of the simulation ($t\!=\!512$) for fixed $t_c\!=\!80$, various values of mode amplitudes $\Gamma_{1,2}$, with noise strength (a) $\langle|\delta\omega|\rangle\!=\!0.001$, and (b) $\langle|\delta\omega|\rangle\!=\!0.01$.
Blue curve with crosses: $\Gamma_1\!=\!0.05$, $\Gamma_2\!=\!0.1$.
Red curve with asterisks: $\Gamma_1\!=\!0.1$, $\Gamma_2\!=\!0.05$.
Black curve with diamonds: $\Gamma_1\!=\!0.05$, $\Gamma_2\!=\!0$.
Green curve with triangles: $\Gamma_1\!=\!0.1$, $\Gamma_2\!=\!0$.}
\end{figure}

One might anticipate that the $n\!=\!2$ mode would couple to a statistically small set of particles in a manner that measurably increases the halo extent beyond that corresponding to the $n\!=\!1$ mode acting alone.
As seen from Fig.~\ref{fig:log_P_r_R_r_2m.ps}, adding the $n\!=\!2$ mode does modify the distribution, though its effect appears to be modest.

\subsubsection{Mixing and Halo Formation}
To visualize how orbits mix, we integrate collections of 1600 test-particle initial conditions clumped into tightly localized regions of phase space.
The integrations are done both without and with noise; the evolution of these collections is depicted in Fig.~\ref{fig:mixing.ps}.
Rows (a)-(c) pertain to the absence of any collective mode, i.e., $\Gamma_1\!=\!\Gamma_2\!=\!0$.
In the absence of noise [row (a)] all mixing is due to a frequency spread across the initial clumps arising from the nonlinear net force.
Separation of the initially localized particle trajectories then proceeds as a power law in time; this is regular phase mixing, i.e., linear Landau damping.
Noise [rows (b) and (c)] influences the effective focusing force acting on the test particles, but this influence generates no significant spreading to large orbital amplitudes.
Turning on the $n\!=\!1$ collective mode changes the situation completely.
With $\Gamma_1\!=\!0.05$, but in the absence of noise [row (d)], the clumps still spread only to a restricted region of phase space; however, the same is clearly not true when noise is included [rows (e) and (f)].
Noise causes a far more efficient mixing.
Even moderately weak noise, e.g., $\langle|\delta\omega|\rangle\!=\!0.01$, thoroughly and exponentially mixes particles, regardless of their starting points, into all regions of the phase space accessed by the beam.
Their exponential separation into global regions of phase space is the principal signature of chaotic mixing~\cite{TE}.
Increasing $\Gamma_1$ further accentuates this chaotic mixing and causes the orbits to fill an even larger phase-space area.

\subsubsection{Circular vs. Radial Orbits}
\label{subsubsec:circular}
Thus far all simulations have pertained to radial test-particle orbits.
Might the results be substantially different for orbits with nonzero angular momentum?
To answer this question, we now consider the other extreme, that for which all test particles are on circular trajectories at $t\!=\!0$; the corresponding values of angular momentum $L$ are assigned according to Eq.~(\ref{eq:L}).
We then solve Eq.~(\ref{eq:motion}) without and with noise in the space-charge tune depression.
This, however, means that we refrain from adding noise to the azimuthal motion.
To do otherwise would vitiate using angular momentum as an integral of the motion and thereby lengthen the computations, all for the `benefit' of incorporating no fundamentally new or important additional phenomenology.

In Fig.~\ref{fig:momentum.ps} results for the halo amplitude $R_H(t)$ [panel (a)] and halo distribution $P(r\!>\!R)$ [panel (b)] are juxtaposed against those pertaining to purely radial orbits.
Although the curves are not identical, neither are they systematically different.
We attribute the differences to statistical fluctuations caused by the random noise included in the simulations.
This finding is interesting in that particles on circular orbits with $r_c(0) \!\geq\!1$ essentially lie outside the influence of the time-dependent potential arising from the collective modes, whereas particles on radial orbits do not.
That halo profiles corresponding to radial versus azimuthal orbits are similar therefore suggests that particles initially in the `core', for which $r(0) \!<\!1$, are the ones that dominate the process of halo formation independent of their initial conditions in velocity space.
By reproducing $R_H(t)$ with the initial radii truncated at $r(0)\!=\!r_c(0)\!=\!1$, we verified that this suggestion is indeed true.

\subsection{Collective Modes vs. Envelope Mismatch}
\label{subsec:mismatch}
The preceding results have illustrated how collective modes, orbital chaoticity, and noise in an envelope-matched beam collaborate to drive an ever-growing halo.
Consider, by contrast, a beam that is envelope-mismatched, i.e., one that exhibits the lowest-order breathing mode~\cite{oconnell,wangler98}.
The corresponding equation of test-particle motion is similar to that governing a beam with a single excited collective mode, except now the `core radius' $R\!=\!R(t)$, i.e., the radius defining the `inside' of the beam, is a function of time.
Specifically, the dimensionless equation governing the core radius is
\begin{equation}
\ddot{R}+R-\frac{\eta^2}{R^3}-\frac{1-\eta^2}{R}~=~0\;,
\label{eq:envelope}
\end{equation}
which then folds into the dimensionless single-particle equation of motion
\begin{eqnarray}
\ddot{r}+r-\frac{1-\eta^2}{R^2}r&=&0\;\;\;\mbox{  for }r\!<\!R(t)\;,\nonumber\\
\ddot{r}+r-\frac{1-\eta^2}{r}&=&0\;\;\;\mbox{  for }r\!\geq\!R(t)\;.
\label{eq:wangler}
\end{eqnarray}
We define the envelope-mismatch parameter $M\!\equiv\!R(t\!=\!0)/R_o$, with $R_o\!=\!1$ denoting the radius of the matched beam.
Results presented for envelope-mismatched beams derive from integrating orbits in the coupled Eqs.~(\ref{eq:envelope})~and~(\ref{eq:wangler}).

To obtain a rough correspondence between the amplitude $\Gamma_1$ of the $n\!=\!1$ collective mode in the envelope-matched beam and the envelope-mismatch parameter $M$, we imagine $M$ to be small.
We then set $R\!\rightarrow\!1\!+\!(M\!-\!1)\cos[\omega_1(\eta^2)t]$, and in turn put $R^{-2}(t) \!\rightarrow\!1\!-\!2(M\!-\!1)\cos[\omega_1(\eta^2)t]$ in Eq.~(\ref{eq:wangler}).
By comparing the end result with Eq.~(\ref{eq:motion}) (with $L$ set to zero) and otherwise neglecting the different definitions of core radii, we then infer $\Gamma_1\!\leftrightarrow\!4(M\!-\!1)^2$.

\begin{figure}
\includegraphics{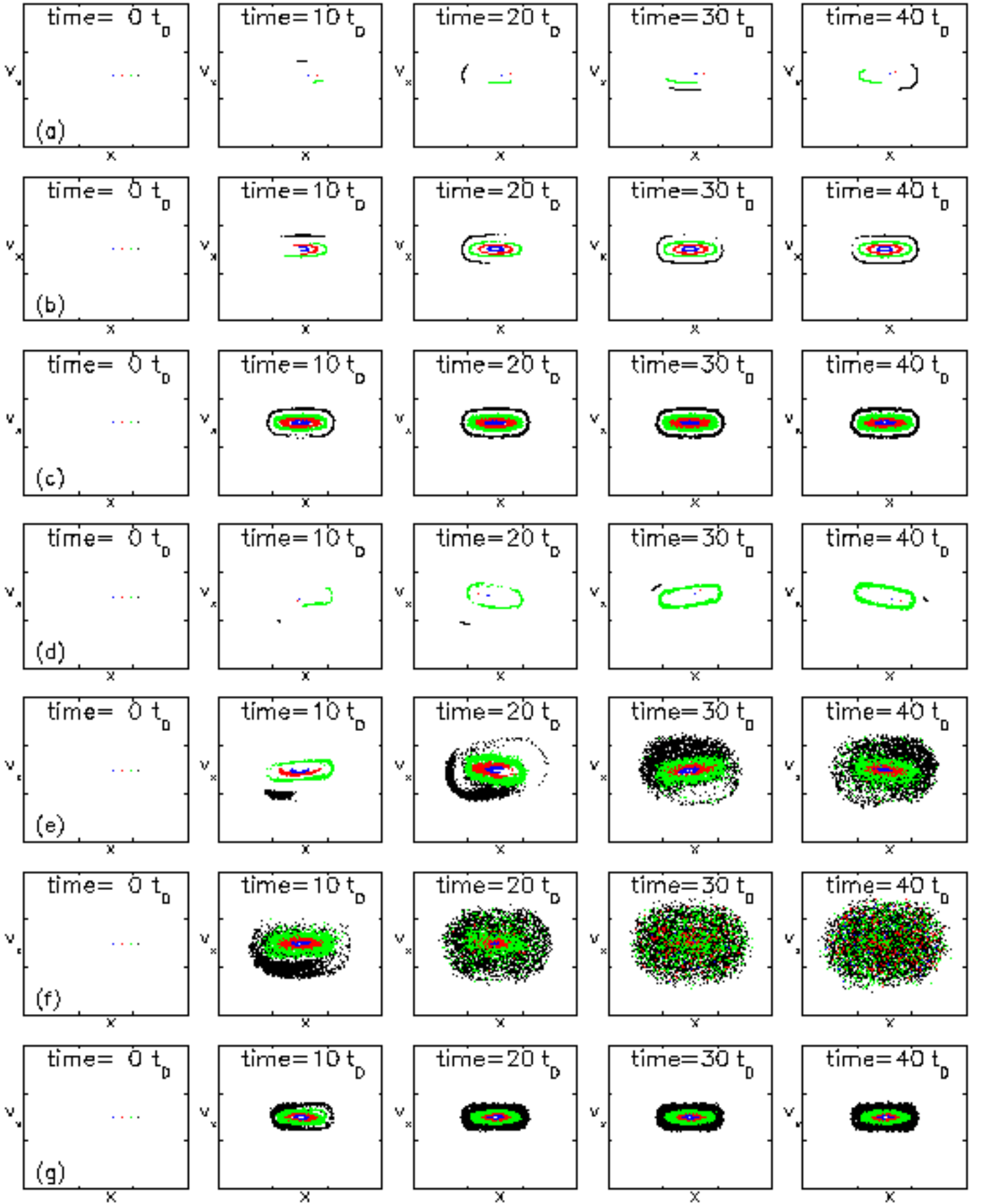}
\caption{\label{fig:mixing.ps}
Evolution of four collections of 1600 test-particle orbits integrated over $t\!=\!512$ DE units ($\sim$60 $t_D$).
The collections start (with zero initial particle velocity) at $x\!=\!0.31\pm 0.002$, $x\!=\!0.70\pm 0.002$, $x\!=\!1.1\pm 0.002$, and $x\!=\!1.41\pm0.002$ with $t_c\!=\!80$, $\Gamma_2\!=\!0$, and:
(a) $\Gamma_1\!=\!0$, $\langle|\delta\omega|\rangle\!=\!0$, 
(b) $\Gamma_1\!=\!0$, $\langle|\delta\omega|\rangle\!=\!0.001$, 
(c) $\Gamma_1\!=\!0$, $\langle|\delta\omega|\rangle\!=\!0.01$, 
(d) $\Gamma_1\!=\!0.05$, $\langle|\delta\omega|\rangle\!=\!0$, 
(e) $\Gamma_1\!=\!0.05$, $\langle|\delta\omega|\rangle\!=\!0.001$, and  
(f) $\Gamma_1\!=\!0.05$, $\langle|\delta\omega|\rangle\!=\!0.01$.
(g) Phase mixing in an envelope-mismatched beam; the mismatch $M\!=\!1.1118$ compares to a mode amplitude $\Gamma_1\!=\!0.05$, and $\langle|\delta\omega|\rangle\!=\!0.01$.}
\end{figure}

\begin{figure}
\includegraphics{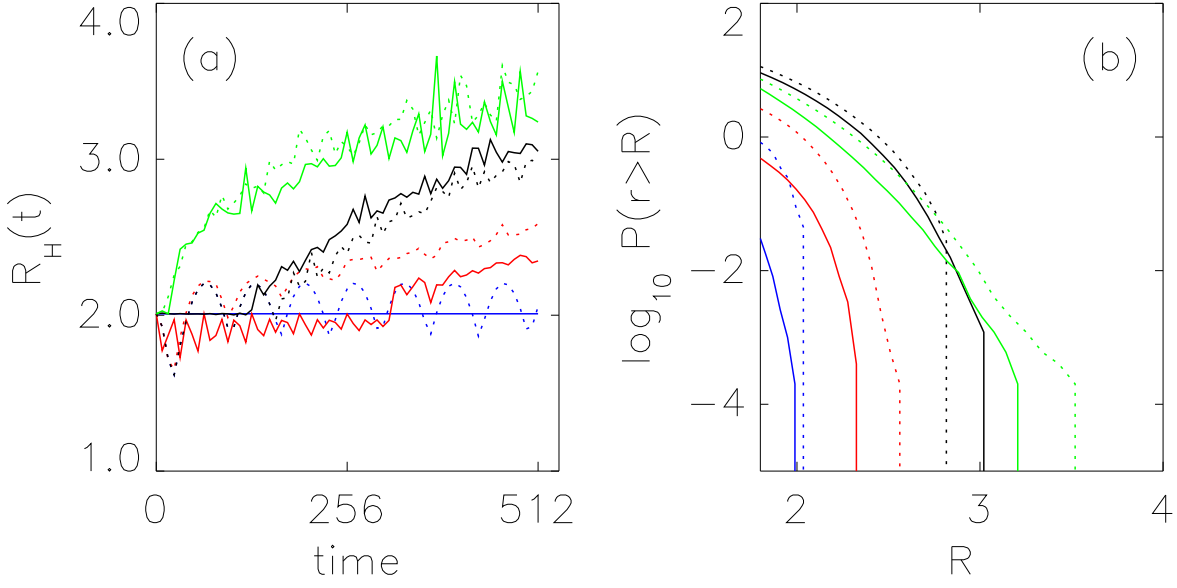}
\caption{\label{fig:momentum.ps}
Halo distributions ($N\!=\!10^6$) for radial orbits (dotted curves) and initially circular orbits (solid curves).
(a) Halo amplitude $R_H(t)$ vs.~$t$ with $\Gamma_1\!=\!0.05$, $\Gamma_2\!=\!0$, and $t_c\!=\!80$.
(b) Percentage $P(r\!>\!R)$ of test particles lying outside radius $R$ at the end of the simulation ($t\!=\!512$).
Blue curves: $\langle|\delta\omega|\rangle\!=\!0$.
Red curves: $\langle|\delta\omega|\rangle\!=\!0.001$.
Black curves: $\langle|\delta\omega|\rangle\!=\!0.01$.
Green curves: $\langle|\delta\omega|\rangle\!=\!0.1$.}
\end{figure}

Now, taking the core radius to be $r\!=\!R(t)$ as pertains to an envelope mismatch rather than $r\!=\!1$ as pertains to collective modes in an envelope-matched beam has a {\it profound} effect on the mixing, hence the halo dynamics.
Figure~\ref{fig:mixing.ps} [row (g)] illustrates the mixing in an envelope-mismatched beam.
Here the mismatch is $M\!=\!1.1118$, for which the counterpart collective-mode amplitude is $\Gamma_1\!=\!0.05$.
Accordingly, the parameters of Fig.~\ref{fig:mixing.ps}(g) are analogous to those of Fig.~\ref{fig:mixing.ps}(f).
What is striking is how comparatively constrained the phase mixing and halo growth turn out to be in the envelope-mismatched beam.
The underlying dynamics are clearly different.
How so?

The answer lies in the Poincar\'{e} surfaces of section (PSS); as SD point out~\cite{strasburg}, these are distinctly different for the two cases.
The PSS for the envelope-mismatched beam exhibits robust, densely packed KAM tori in the region of phase space exterior to the beam.
This is true even if the mismatch is large.
These tori inhibit the particles from gaining significant energy and reaching large amplitudes.
The $n\!=\!1$ collective mode in the envelope-matched beam, by contrast, weakens the tori in the vicinity of the beam edge $r\!=\!1$.
As a consequence, particles are then freer to move; they can stochastically and rapidly explore a large region of phase space.
SD point out that the orbital amplitudes of those particles can rapidly increase as the amplitude of the collective mode is raised, whereas for the envelope-mismatched beam, test particles gain negligible energy as the mismatch is raised.
SD's findings pertain to zero noise; we find nonzero noise substantially magnifies them by further weakening and/or breaking the tori.
When the KAM tori are broken, a series of small, successive kicks can much more easily push a particle to ever increasing radii.
Moreover, the associated time scale is short; significant extended halo can form in just a few dynamical times, i.e., orbital periods.

\subsection{Noise-Induced Breakdown of Tori}
\label{subsec:tori}
\begin{figure}
\includegraphics{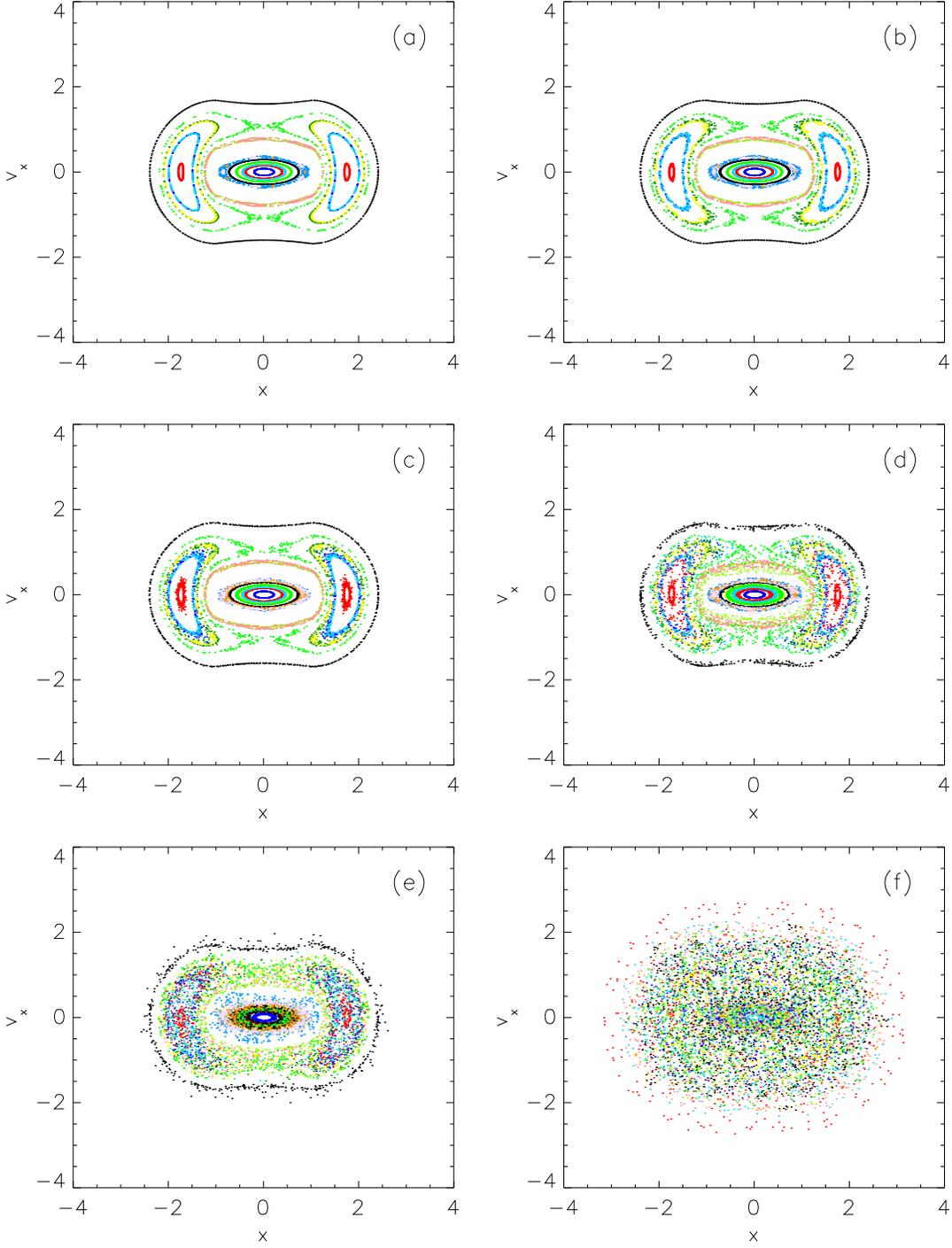}
\caption{\label{fig:psection_G_0.05.ps}
Poincar\'{e} sections for a set of 18 representative initial conditions integrated over $t\!=\!2048$ DE units ($\sim$250 $t_D$) with various noise strengths, and with $t_c\!=\!80$, $\Gamma_1\!=\!0.05$, and $\Gamma_2\!=\!0$:
(a) $\langle|\delta\omega|\rangle\!=\!0$, 
(b) $\langle|\delta\omega|\rangle\!=\!10^{-6}$, 
(c) $\langle|\delta\omega|\rangle\!=\!10^{-5}$, 
(d) $\langle|\delta\omega|\rangle\!=\!10^{-4}$, 
(e) $\langle|\delta\omega|\rangle\!=\!5\times 10^{-4}$, and  
(f) $\langle|\delta\omega|\rangle\!=\!10^{-3}$.}
\end{figure}

To visualize noise-induced disintegration of tori with consequent halo formation, we plot the Poincar\'{e} sections of 18 test-particle orbits having initial conditions that collectively represent the whole of configuration space.
We take only the $n\!=\!1$ mode to be excited, with amplitude $\Gamma_1\!=\!0.05$, and integrate the 18 trajectories for a total time $t\!=\!2048$ (about 250 $t_D$).
We do a series of these experiments, starting with a noise strength $\langle|\delta\omega|\rangle\!=\!10^{-6}$ and successively increasing it to $10^{-5}$, $10^{-4}$, $5\times 10^{-4}$, and $10^{-3}$; for every experiment we set $t_c\!=\!80$.
We record the positions and velocities of the particles at every period $T\!=\!2\pi/\omega_1$.
For each experiment the respective PSS is shown in Fig.~\ref{fig:psection_G_0.05.ps}; different colors denote different orbits.

The Poincar\'{e} sections clarify the underlying microscopic dynamics.
As the noise strength is raised, `internal' (lower-energy) tori are clearly the first, and thus the easiest, to break.
With stronger noise the outermost tori break as well.
Note that the strongest noise considered is only a 0.1\% fluctuation of the mode frequency, and yet this noise breaks all of the tori [panel (f)].
It is important to remember that these plots lack statistical significance since only 18 orbits are represented.
In a statistically important, i.e., much larger, sample some number of particles may conceivably break through the outer tori even with very small noise.
What the plots suggest is that this number should increase as the noise strength increases, in keeping with what one would expect intuitively.
The `disintegration time scale' (delineating the onset of halo formation) is as indicated in the plots of $R_H$ discussed earlier, and these plots were developed with meaningful statistics, i.e., with $10^6$ test particles.

\begin{figure}
\includegraphics{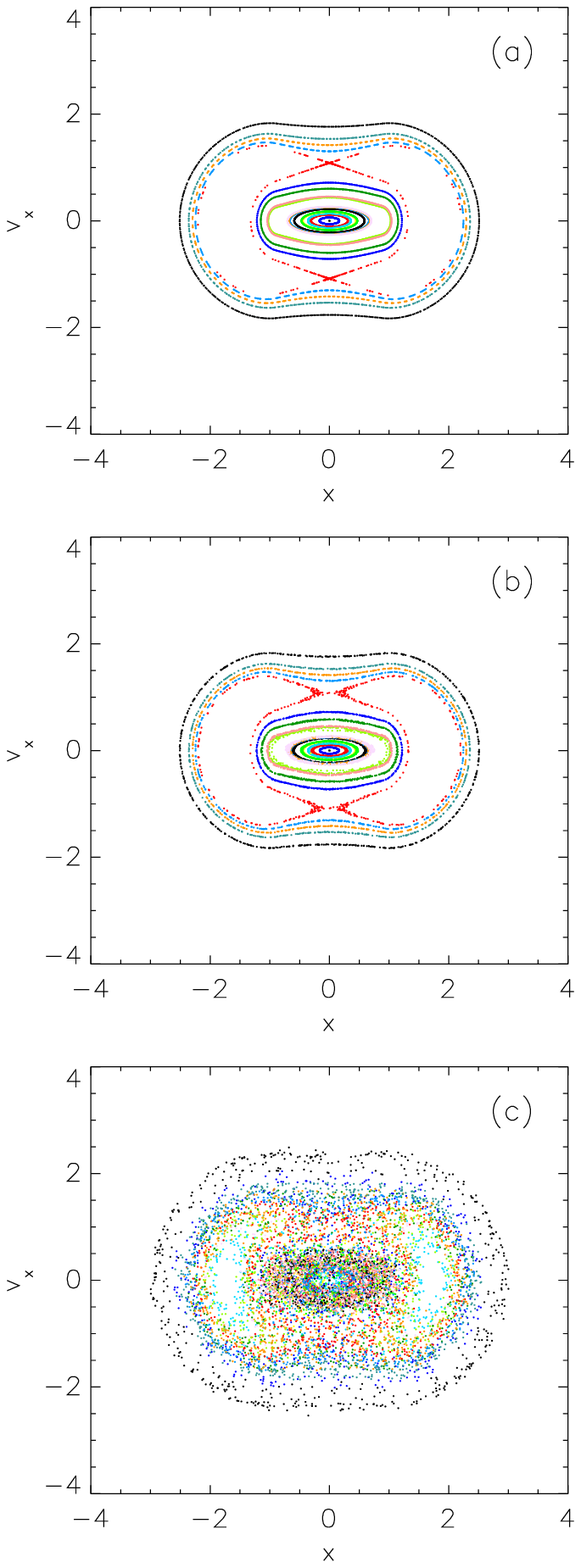}
\caption{\label{fig:wsection_G_0.05.ps}
Poincar\'{e} sections for a set of 18 representative initial conditions in an envelope-mismatched beam integrated over $t\!=\!2048$ DE units ($\sim$250 $t_D$) with various noise strengths, and with $M\!=\!1.1118$:
(a) $\langle|\delta\omega|\rangle\!=\!0$, 
(b) $\langle|\delta\omega|\rangle\!=\!10^{-3}$, and 
(c) $\langle|\delta\omega|\rangle\!=\!10^{-1}$.}
\end{figure}

We now repeat the same investigation for the corresponding envelope-mismatched beam, i.e., $M\!=\!1\!+\!\Gamma^{1/2}_1/2\!=\!1.1118$.
Here, we record the positions and velocities of the particles at every period at which the core has its minimum radius; the results appear in Fig.~\ref{fig:wsection_G_0.05.ps}.
Panel (a) depicts the PSS with zero noise, whereas panel (b) depicts the PSS with noise having the same parameters ($\langle|\delta\omega|\rangle\!=\!10^{-3}$ and $t_c\!=\!80$) as in Fig.~\ref{fig:psection_G_0.05.ps}(f).
Although with this noise the KAM tori become slightly fuzzier, they are not yet broken, and this stands in stark contrast to the situation wherein a collective mode of similar amplitude is active in an envelope-matched beam.
Even when the noise has very large amplitude [$\langle|\delta\omega|\rangle\!=\!10^{-1}$ in Fig.~\ref{fig:wsection_G_0.05.ps}(c)], the beam boundary in phase space appears still to be sharply defined, although the tori have now obviously broken to a certain degree, particularly in the beam's interior.
Hence, although the tori of the envelope-mismatched beam are seen to be robust, they are not indestructible; sufficiently strong noise will eventually break them.

\section{Discussion and Conclusions}
\label{sec:discussion}
The foregoing results illustrate that collective modes can have a critical impact on halo formation by destabilizing the phase space near the beam boundary.
Particles then are freer to roam by interacting with the collective modes, extracting energy from them, and thereby populating a halo.
By keeping a statistically small number of particles in phase with collective modes, colored noise contributes toward not only populating the halo, but also expanding its extent, and it does so rapidly.

One might reasonably question, because they are unrealistic, whether density discontinuities inherent to collective modes in the warm-fluid KV beam might vitiate our findings by imposing correspondingly unrealistic dynamics.
The answer would seem to be no because we showed in earlier work (Ref.~\cite{bs03}) that adding noise to a perturbed thermal-equilibrium beam, a beam that is devoid of discontinuities and wherein the perturbation mimics the presence of a global collective mode, yields orbital amplitudes entirely consistent with those computed herein.

These matters are of practical importance to the evolution of real beams.
Transitions in an accelerator will give rise to various mismatches that move the beam away from equilibrium.
Subsequent charge redistribution will trigger a hierarchy of collective modes.
Unavoidable irregularities in the beamline will impose a spectrum of colored noise that adds self-consistently to the time-dependent potential associated with the collective modes.
Consequently, the phenomenology that we uncovered will arise, as will the consequential growth of the beam's phase space in general, and beam halo in particular.
Accounting for these details therefore becomes {\it imperative}, particularly in regard to designing accelerators for the production of high-average-current beams.

Although by working with the warm-fluid model of a beam we have endeavored toward a treatment that is realistic, yet still generic, our treatment nevertheless retains some shortcomings that need to be rectified in future work.
These include the following: (1) The distribution of collective modes will evolve in a real beam~\cite{bohn93}; modes will tend to dissipate in conjunction with the redistribution of the free energy they contain, a dynamic that we have neglected.
However, the time scales over which large-scale collective modes dissipate are not yet well quantified, and evidence from numerical simulations suggest they may persist for hundreds of dynamical times~\cite{haber04}.
To the extent this proves true, our analysis reveals the attendant impact on halo formation.
(2) A real beam contains no test particles; all of the particles interact with one another.
(3) A real accelerator will present a spectrum of colored noise, i.e., a distribution of noise parameters, in keeping with the actual hardware and field irregularities.
The totality of this phenomenology can be incorporated only by way of careful self-consistent $N$-body simulations that reflect both accurate boundary conditions and statistically accurate initial conditions, as well as faithfully reproduce the hierarchies of spatial and temporal scales intrinsic to the evolving beam.

We have endeavored to show clearly and convincingly that details can be important to the evolution of a charged-particle beam under the influence of space charge, in that they can make a substantial impact on the macroscopic evolution of its phase-space distribution.
Accordingly, these details merit careful study.
A seemingly probable outcome would be that the proper way to picture generically a nonequilibrium beam subject to self-forces is in terms of an increasingly well-mixed and continually growing phase space as opposed to a phase space in which tori largely partition, and hence constrain, the motion of the constituent particles.
This is especially true considering that the results herein pertain to 1.5-dimensional beams (the half dimension corresponding to time), whereas real beams are higher-dimensional systems, and thus their phase spaces are inherently less hospitable to barriers in the form of tori and cantori.

In a paper that appeared subsequent to the submission of our manuscript, Gerigk shows that modest statistical errors in the focusing gradient generate continually growing halo~\cite{gerigk04}.
His initial study centers on a cylindrical beam that, in the absence of these errors, is in equilibrium, e.g., it is both rms-matched and envelope-matched.
He finds that the focusing errors excite oscillations in this matched beam that then transfer energy to single particles.
He also demonstrates that the same phenomenology applies in a more realistic three-dimensional, i.e., bunched, beam.
Inasmuch as Gerigk was unaware of our earlier paper (Ref.~\cite{bs03})~\cite{gerigkcom}, and we were likewise unaware of his related activity, through complementary investigations we have all independently arrived at the same conclusion: noise constitutes a continual source of enhanced halo production.

\begin{acknowledgments}
This work was supported by the Department of Education under Grant No. P116Z010035, and by the Department of Energy under Grant No. DE-FG02-04ER41323.
\end{acknowledgments}

\clearpage

\clearpage

\clearpage

\clearpage

\clearpage

\clearpage

\clearpage

\clearpage

\clearpage

\clearpage


\begin{thebibliography}{99}
\bibitem{bs03}C. L. Bohn and I. V. Sideris, Phys. Rev. Lett. {\bf 91}, 264801 (2003).
\bibitem{chen91}Y.-J. Chen et al., in Proceedings of the 1991 Particle Accelerator Conference, edited by L. Lazema and J. Chew (IEEE, Piscataway, NJ, 1991), p. 3100.
\bibitem{gluckstern94}R. L. Gluckstern, Phys. Rev. Lett. {\bf 73}, 1247 (1994).
\bibitem{jameson96}R. Jameson, Fus. Eng. Des. {\bf 32-33}, 149 (1996).
\bibitem{lund98}S. Lund and R. C. Davidson, Phys. Plasmas {\bf 5}, 3028 (1998).
\bibitem{llbook}A. J. Lichtenberg and M. A. Lieberman, {\em Regular and Stochastic Motion} (Springer-Verlag, Berlin, 1983).
\bibitem{strasburg}S. Strasburg and R. C. Davidson, Phys. Rev. E {\bf 61}, 5753 (2000).
\bibitem{kandrup03}H. E. Kandrup, I. V. Sideris, and C. L. Bohn, Phys. Rev. ST Accel. Beams {\bf 7}, 014202 (2003).
\bibitem{qiang01}J. Qiang {\it et al.}, Nucl. Instrum. Methods Phys. Res. Sect. A {\bf 457}, 1 (2001).
\bibitem{van} N. G. van Kampen, {\em Stochastic Processes in Physics and Chemistry} (North Holland, Amsterdam, 1981).
\bibitem{pogorelov99}I.V. Pogorelov and H.E. Kandrup, Phys. Rev. E {\bf 60}, 1567 (1999).
\bibitem{reiser}M. Reiser, {\em Theory and Design of Charged Particle Beams} (Wiley, NY, 1994), cf. Section 5.4.
\bibitem{brown95}N. Brown and M. Reiser, Phys. Plasmas {\bf 2}, 965 (1995).
\bibitem{TE}C. L. Bohn and I. V. Sideris, Phys. Rev. ST Accel. Beams {\bf 6}, 034203 (2003).
\bibitem{qiang02}J. Qiang, et al., Phys. Rev. ST Accel. Beams {\bf 5}, 124201 (2002).
\bibitem{NR}W. H. Press, S. A. Teukolsky, W. T. Vetterling, and B. P. Flannery, {\em Numerical Recipes in C} (Cambridge University Press, Cambridge, 1995).
\bibitem{sns}Spallation Neutron Source Report No. 100000000-PL0001-R10 (unpublished).
\bibitem{oconnell}J. S. O'Connell, T. P. Wangler, R. S. Mills, and K. R. Crandell, in Proceedings of the 1993 Particle Accelerator Conference, edited by S.T. Corneliussen (IEEE, Piscataway, NJ, 1993), p. 3657.
\bibitem{wangler98}T. P. Wangler, K. R. Crandall, R. Ryne, and T. S. Wang, Phys. Rev. ST Accel. Beams {\bf 1}, 084201 (1998).
\bibitem{bohn93}C. L. Bohn, Phys. Rev. Lett. {\bf 70}, 932 (1993).
\bibitem{haber04}I. Haber, et al., Nucl. Instrum. Methods {\bf A519}, 396 (2004).
\bibitem{gerigk04}F. Gerigk, Phys. Rev. ST Accel. Beams {\bf 7}, 064202 (2004).
\bibitem{gerigkcom}F. Gerigk, private communication.
\end{thebibliography}
\end{document}